\documentclass[twocolumn,preprintnumbers,amsmath,amssymb]{revtex4}
\usepackage{dcolumn}
\usepackage{bm}
\usepackage{graphicx,subfigure,xcolor}
\usepackage{braket}
\usepackage[normalem]{ulem}

\DeclareMathAlphabet\mathbfcal{OMS}{cmsy}{b}{n}

\usepackage{comment}
\usepackage{xcolor}
\usepackage{amsmath}
\usepackage{amssymb}
\usepackage{eucal}
\usepackage{mathrsfs}
\usepackage{amsthm}
\usepackage{epstopdf}
\usepackage{textcomp}
\usepackage{braket}
\usepackage{pdfpages}
\usepackage{soul}

\def\br{{\bf r}}

\def\bk{{\bf k}}

\begin{document}

\title{Dispersion interaction between two hydrogen atoms in a static electric field}

\author{Giuseppe Fiscelli}
\email{giuseppe.fiscelli@unipa.it}

\author{Lucia Rizzuto}
\email{lucia.rizzuto@unipa.it}

\author{Roberto Passante}
\email{roberto.passante@unipa.it}

\affiliation{Dipartimento di Fisica e Chimica - Emilio Segr\`{e}, Universit\`{a} degli Studi di Palermo, Via Archirafi 36, I-90123 Palermo, Italia \\
and INFN, Laboratori Nazionali del Sud, I-95123 Catania, Italy}

\begin{abstract}
We consider the dispersion interaction between two ground-state hydrogen atoms, interacting with the quantum electromagnetic field in the vacuum state, in the presence of an external static electric field, both in the nonretarded and in the retarded Casimir-Polder regime. We show that the presence of the external field strongly modifies the dispersion interaction between the atoms, changing its space dependence. Moreover, we find that, for specific geometrical configurations of the two atoms with respect to the external field and/or the relative orientation of the fields acting on the two atoms, it is possible to change the character of the dispersion force, turning it from attractive to repulsive, or even make it vanishing. This new finding clearly shows the possibility to control and tailor interatomic dispersion interactions through external actions. By a numerical estimate of the field-modified interaction, we show that at typical interatomic distances the change of the interaction's strength can match or even outmatch the unperturbed interaction; this can be obtained for values of the external field that can be currently achieved in the laboratory, and sufficiently weak to be taken into account perturbatively.
\end{abstract}

\maketitle

\textit{Introduction.}
Radiation-mediated interactions, such as Casimir-Polder and van der Waals forces, are long-range interactions between atoms, molecules or macroscopic bodies, related to the zero-point fluctuations of the quantum electromagnetic field  \cite{CasimirPolder48,CP98,CPP95,Passante18,Milonni94,Salam10,EGJ13,AS87,Bartolo15,Donaire16,Adhikari18}. These ubiquitous interactions have great relevance in many areas of physics, and applications in biology \cite{Preto15,CS14}, chemistry \cite{Galego19} and nanotechnologies \cite{Chan01,Capasso11,Decca11}.
A striking property is the possibility to control and tailor them through the external environment, for example through cavities or walls \cite{PT82,Cho96,Spagnolo06,Safari06,Ellingsen10}, waveguides \cite{Shahmoon13,Shahmoon14,Haack15,Fiscelli17,Fiscelli18}, or photonic crystals \cite{Incardone14,Notararigo18}. Of special interest is to investigate the possibility of making  Casimir interactions repulsive, in particular for applications in nano- and micro-electromechanical systems (NEMS and MEMS), for which Casimir interactions between their various parts, at sub-micrometrical separation, can be relevant \cite{Chan01}; a repulsive Casimir interaction could help to prevent stiction and consequent failure of these devices \cite{Bordag09,Bostrom12}.

In recent years, the possibility of manipulating van der Waals interactions through external optical fields has been investigated \cite{Thiru80,Milonni96,Sherkunov09}; it was shown that a near-resonant electromagnetic field can enhance the intensity of the atom-surface Casimir-Polder force \cite{Perrealult08}, and that, in the presence of an external quantum field, the Casimir-Polder interaction between a ground-state atom and a nondispersive surface has features similar to the Casimir-Polder potential for excited atoms \cite{Haug18}, leading to the possibility of a repulsive force \cite{Fuchs18}. Furthermore, by inducing an atomic rotating dipole by polarized light, possibility of a lateral or repulsive force has been predicted \cite{Barcellona18}. The effect of a static electric field on the dipole-dipole near-zone van der Waals interaction in ultracold Rydberg gases has been recently studied \cite{Jahangiri19,You98}.

In this Letter we consider a different situation, that is, the dispersion interaction between two ground-state hydrogen atoms, in both the nonretarded and the retarded (Casimir-Polder) regime, subjected to an external static electric field. Due to the presence of this field, the atoms acquire spatially correlated dipole moments; we find that this strongly affects the atom-atom dispersion interaction, modifying its distance dependence, even for a low intensity of the external field that can be easily obtained in the laboratory, and such that can be treated perturbatively.
In the geometrical configurations which we consider in detail, we show that the distance dependence of the dispersion interaction changes, decreasing slower with the distance (as $r^{-3}$ and $r^{-4}$ in the near and far zone respectively, compared to $r^{-6}$ and $r^{-7}$ for unperturbed atoms). Also, we find that we can modify the magnitude of the dispersion interaction, its attractive/repulsive character, through the external field, or causing it to vanish, exploiting a field strength achievable in current experimental setups. A clear physical interpretation of these findings is discussed, as well as a numerical estimate of the effect, and its observability and relevance. These results can be relevant for a direct measurement of dispersion interactions between atoms \cite{Beguin13,Przybytek12}.

\textit{The field-modified dispersion interaction.}
We consider two hydrogen atoms, $A$ and $B$, interacting with the quantum electromagnetic field in the vacuum state, subjected to the external static and uniform (over the atomic dimensions) classical electric fields
$\mathbfcal{E}$ and $\mathbfcal{E}'$, respectively.
We assume that atom $A$ is at ${\bf r}=0$, and ${\bf r}$ is the separation vector between the two atoms.
The unperturbed ground state is $\lvert \phi_{100}^A, \phi_{100}^B \rangle \mid 0_{\bk \lambda} \rangle$, where $\lvert \phi_{n\ell m} \rangle$ is an atomic state with quantum numbers $n,\ell ,m$, and  $\lvert 0_{\bk \lambda} \rangle$ is the electromagnetic vacuum state. This state is corrected by both the external electric field and the quantum radiation field.

The Hamiltonian of our system is
\begin{equation}
\label{Hamiltonian}
H=H_0 + V_A + V_B + H_I^A + H_I^B ,
\end{equation}
where $H_0=H_A+H_B+H_F$ is the unperturbed Hamiltonian, sum of the atomic and the field Hamiltonians; $V_{A(B)}$ and $H_I^{A(B)}$ are, respectively, the interaction Hamiltonian of atom $A(B)$ with the external field and the radiation field. In the multipolar coupling scheme and within the dipole approximation, they are
\begin{eqnarray}
V_A= -\bm{\mu}_A \cdot \mathbfcal{E} ; \, \, \, \, V_B = -\bm{\mu}_B \cdot \mathbfcal{E}' , \\
H_I^A = -\frac 1{\epsilon_0}\bm{\mu}_A\cdot\mathbf{D}_\perp (\mathbf{0}) ; \, \, \, \, H_I^B = -\frac 1{\epsilon_0}\bm{\mu}_B\cdot\mathbf{D}_\perp (\mathbf{r}),
\label{IntHam}
\end{eqnarray}
where $\bm{\mu}_{A(B)}$ is the dipole moment operator of atom $A(B)$, proportional to the electric charge $q$, and
\begin{equation}\label{elfield}
\mathbf{D}_\perp (\br ) = i\sum_{\bk \lambda} \sqrt{\frac {\hbar c k\epsilon_0}{2V}} \hat{e}_{\bk \lambda} \left( a_{\bk \lambda} e^{i\bk \cdot \br} - a_{\bk \lambda}^\dagger e^{-i\bk \cdot \br} \right)
\end{equation}
is the transverse displacement field operator \cite{CP98,Passante18}. In \eqref{elfield}, $V$ is the quantization volume, $\epsilon_0$ is the vacuum dielectric constant, $\hat{e}_{\bk \lambda}$ are real polarization unit vectors ($\lambda =1,2$), and $a_{\bk \lambda},a_{\bk \lambda}^\dagger$ are annihilation and creation operators satisfying bosonic commutation rules.

In the absence of the external field, the dispersion interaction between the two atoms is at fourth order in the coupling with the radiation field \cite{CasimirPolder48,CPP95,CP98,Passante18}: vacuum fluctuations induce and correlate dipole moments in the atoms (this accounts for two orders in the atom-field coupling), and, afterwards, the correlated dipoles yield an interaction energy (two more orders in the coupling constant) \cite{PT93,Passante18}. This is equivalent to the exchange of a pair of virtual photons. In the present case, for the dispersion interaction component related to the external field: i) the external fields polarize and correlate the atoms, which causes them to acquire a permanent dipole moment (first order in the coupling with the external field, for each atom); ii) these dipoles interact through the transverse field (one virtual-photon exchange, thus a second-order process in the coupling with the transverse field). Although the complete interaction energy is proportional to $q^4$, a factor $q^2$ is related to the interaction with the external fields, and proportional to their intensity, and another factor $q^2$ is related to the interaction with the transverse field. The interatomic potential involves the exchange of just one virtual photon, similarly to the atom-wall Casimir-Polder interaction \cite{CasimirPolder48,Buhmann1} or the entangled-atoms resonance interaction \cite{CP98,Incardone14}.

We first obtain the ground state corrected up to second order in the external field; then, in order to obtain the dispersion interaction, we evaluate the distance-dependent part of the second-order energy shift due to the interaction with the transverse field. For simplicity, we assume $\mathbfcal{E}$ and $\mathbfcal{E}'$ along the $z$-axis (positive or negative), and include only contributions from $n=2$ atomic intermediate states (inclusion of higher $n$ atomic states does not change the qualitative features of our results). The second-order ground state, eigenstate of $H_0+V_A+V_B$, in the presence of the static fields, is
\begin{eqnarray}
\label{corrgroundstate}
\mid \psi \rangle &=& \Big\{ \left[ 1- \gamma^2 (\mathcal{E}^2 + {\mathcal{E}'}^2 )\right] \lvert \phi_{100}^A, \phi_{100}^B \rangle
\nonumber \\
&-& \sqrt{2} \gamma \left( \mathcal{E} \lvert \phi_{210}^A, \phi_{100}^B \rangle  +  \mathcal{E}' \lvert \phi_{100}^A, \phi_{210}^B \rangle \right)
\nonumber \\
&+&  \gamma^2 \Big[ 2 \mathcal{E} \mathcal{E}'  \lvert \phi_{210}^A, \phi_{210}^B \rangle - \frac 1{\sqrt{2}} \left( \frac 32 \right)^6
\nonumber \\
&\times& \! \! \! \left( \mathcal{E} ^2  \lvert \phi_{200}^A, \phi_{100}^B \rangle
+  {\mathcal{E}'}^2 \lvert \phi_{100}^A, \phi_{200}^B \rangle \right) \Big] \Big\} \mid 0_{\bk \lambda} \rangle ,
\end{eqnarray}
where $\gamma =2^9qa_0/(3^6E_1)$, $E_1$ and $E_2=E_1/4$ are, respectively, the unperturbed energies of the ground and the first excited level of the hydrogen atom, and $a_0$ the Bohr radius.

The second-order energy correction of the state \eqref{corrgroundstate}, due to the interaction \eqref{IntHam} with the radiation field, is

\begin{equation}
\Delta E = \sum_I \frac {\langle \psi \lvert (H_I^A+H_I^B)\rvert I \rangle \langle I \lvert (H_I^A+H_I^B) \rvert \psi \rangle}{E_\psi-E_I},
\label{enshift}
\end{equation}
where $E_\psi$ is the energy of the state \eqref{corrgroundstate}, including the Stark shift, and $\lvert I \rangle$ are the other eigenstates of $H_0+V_A+V_B$, with energy $E_I$, obtained by degenerate-state perturbation theory (the $n=2$ level has a fourfold degeneracy).

We need to evaluate only the part of the shift \eqref{enshift} that depends on the interatomic distance (self-energies do not contribute to the interaction energy \cite{CP98}), and thus only terms containing both $H_I^A$ and $H_I^B$ are relevant; this introduces a factor proportional to $q^2$, due to the atomic dipole moments in \eqref{IntHam}. We now determine which intermediate states
$\lvert I \rangle = \lvert \tilde{I} \rangle \lvert n_{\bk \lambda} \rangle$
contribute to the field-modified fourth-order energy shift. The field part  of these intermediate states consists of one-photon states $\lvert 1_{\bk \lambda} \rangle$, while
the atomic part consists of the eigenstates $\lvert \tilde{I} \rangle$ of $H_A + H_B + V_A + V_B$, which are obtained by degenerate-state perturbation theory, and written in the form
$\lvert \tilde{I} \rangle = \lvert \tilde{I}_0\rangle + \lvert \tilde{I}_1\rangle + \lvert \tilde{I}_2\rangle$, where the subscript $0,1,2$ indicates the perturbative order in $V_A + V_B$. The zeroth-order terms are
\begin{equation}
\label{Intermediate}
\lvert \tilde{I}_0^{\pm , \pm}\rangle =  \frac 12 \left( \lvert \phi_{210}^A \rangle \pm \lvert \phi_{200}^A \rangle \right)
\left( \lvert \phi_{210}^B \rangle \pm \lvert \phi_{200}^B \rangle \right) .
\end{equation}
The four intermediate states \eqref{Intermediate} have the same unperturbed energy, and give a nonvanishing matrix element in \eqref{enshift} due to the terms linear in $\mathcal{E}$ and $\mathcal{E}'$ in \eqref{corrgroundstate}; for the $\br$-dependent contributions, we have
\begin{align}
\label{Intmatrelem}
&\sum_{m,n=+,-} \langle \psi \lvert H_I^A \rvert \tilde{I}_0^{mn} 1_{\bk \lambda} \rangle  \langle \tilde{I}_0^{mn} 1_{\bk \lambda} \lvert H_I^B \rvert \psi \rangle + (A \leftrightarrow B)
\nonumber \\
&\ = 8 \gamma^2 \mathcal{E} \mathcal{E}' \frac {\hbar ck}{2V\epsilon_0} (\hat{e}_{\bk \lambda} \cdot \bm{\mu}_B^{ge}) (\hat{e}_{\bk \lambda} \cdot \bm{\mu}_A^{eg}) e^{i\bk \cdot \br} ,
\end{align}
where $\bm{\mu}_{A(B)}^{eg} = \langle \phi_{210} \lvert \bm{\mu}_{A(B)} \rvert \phi_{100} \rangle = 2^{15/2} 3^{-5} q a_0 \hat{z}$ is oriented along $z$.
The energy denominator in \eqref{enshift}, relative to intermediate states \eqref{Intermediate}, is $2(E_2-E_1) +\hbar ck = 3\lvert E_1 \rvert /2 +\hbar ck = \hbar c (k_0 +k)$, where $k_0=2 \lvert E_2-E_1 \rvert /\hbar c$.
Terms from first- and second-order intermediate states, $\lvert \tilde{I}_1\rangle$ and $\lvert \tilde{I}_2\rangle$, only yield higher-order contributions to the energy shift, and can be neglected. Also, because the numerator in \eqref{enshift} is at fourth order in $q$, the Stark energy shifts in $E_\psi$ and $E_I$ can be neglected.
Using \eqref{Intmatrelem} in \eqref{enshift}, finally yields, for the field-assisted distance-dependent energy shift,
\begin{equation}
\label{enshift1}
\Delta E (r) = -\frac {4\gamma^2 \mathcal{E} \mathcal{E}'}{V \epsilon_0}  \sum_{\bk \lambda} (\hat{e}_{\bk \lambda} \cdot \bm{\mu}_B^{ge}) (\hat{e}_{\bk \lambda} \cdot \bm{\mu}_A^{eg})
\frac {ke^{i\bk \cdot \br} }{k+k_0} .
\end{equation}

After performing the polarization sum, $\sum_\lambda (\hat{e}_{\bk \lambda})_i (\hat{e}_{\bk \lambda})_j = \delta_{ij}-\hat{k_i} \hat{k}_j$, taking the continuum limit, $\sum_\bk \rightarrow V/(2\pi )^3 \int dk k^2 \int d\Omega$, and performing algebraic calculations involving angular and frequency integrations, Eq. \eqref{enshift1} gives
\begin{eqnarray}\label{enshift2}
\Delta E (r) &=& -\frac {2\gamma^2}{\pi^2 \epsilon_0} \mathcal{E} \mathcal{E}' (\mu_A^{eg})_i (\mu_B^{ge})_j
\nonumber \\
&\ & \times \left( -\nabla^2 \delta_{ij} + \nabla_i \nabla_j \right) \frac 1r f(k_0r) ,
\end{eqnarray}
where the Einstein convention of repeated indices has been used, the differential operators act on the variable $\br$, and we introduce the auxiliary functions
\begin{eqnarray}
f(x) &=&{\rm Ci}(x) \sin (x) + \left( \frac \pi 2 - {\rm Si} (x) \right) \cos (x), \\
g(x) &=& - {\rm Ci} (x) \cos (x) + \left( \frac \pi 2 - {\rm Si} (x) \right) \sin (x),
\end{eqnarray}
with $\rm{Si} (x)$ and $\rm{Ci} (x)$ being the sine and cosine integral functions \cite{NIST}.

Expression \eqref{enshift2} is valid for a generic geometrical configuration of the atoms with respect to the external fields. It should be added to the usual unperturbed dispersion interaction for ground-state atoms ($\Delta E_{vdW}(r)$, given by our ulterior Eq. \eqref{vdW}). It is a second-order quantity in the coupling with the radiation field, involving one virtual photon exchange; it also contains a second-order coupling with the external fields. It is thus an overall fourth-order quantity in $q$ (each dipole moment $\mu$ brings one power of $q$). It is proportional to $\mathcal{E}$ and $\mathcal{E}'$, allowing an external control of its intensity and sign.

We now focus on two relevant geometrical configurations: atoms aligned in a direction perpendicular or parallel to the direction of the external field.

For atoms aligned perpendicularly to the external field, Eq. \eqref{enshift2} yields
\begin{equation}\label{enshiftperp1}
\Delta E_\perp (r) = \frac{\beta \mathcal{E} \mathcal{E}'}r \left[ f(k_0r)\left( \frac 1{(k_0r)^2}-1\right) + \frac {g(k_0r)}{k_0r} +\frac 1{k_0r} \right] ,
\end{equation}
where we have defined
\begin{equation}\label{constant}
\beta = \frac{2\gamma^2 k_0^2\mu_A^{eg} \mu_B^{ge}}{\pi^2 \epsilon_0} = \frac 1{4\pi \epsilon_0} \frac {2^{34}q^4a_0^4}{3^{20}\pi \hbar^2 c^2} = \frac {9k_0^2}{4\pi^2 \epsilon_0}\alpha^2 ,
\end{equation}
where $E_1 = -\hbar^2/(2ma_0^2)$, $\alpha = 2\mu^2 /[3(E_2-E_1)]$ is the static polarizability of the atoms, and ${\bm \mu}^{eg}$, $k_0$ have been defined before.

Using the asymptotic expansions of the auxiliary functions $f(x)$ and $g(x)$ \cite{NIST}, this expression can be approximated in the near ($k_0r \ll 1$) and far ($k_0r \gg 1$) zone, yielding
\begin{equation}\label{enshiftperp2}
\Delta E_\perp (r) \simeq \beta \mathcal{E} \mathcal{E}' \frac 1{k_0^2} \times
\left\{
\begin{array}{cc}
& \frac \pi{2r^3} \, \, \, \, \, \mbox{for $k_0r \ll 1$} \\
& \\
& \frac 4{k_0r^4} \, \, \, \,  \mbox{for $k_0r \gg 1$} .
\end{array}
\right.
\end{equation}

Similarly, for atoms aligned in the same direction of the external fields,
\begin{equation}\label{enshiftpar1}
\Delta E_\parallel (r) = -2\beta \mathcal{E} \mathcal{E}' \frac 1r \left[ \frac {f(k_0r)}{(k_0r)^2} + \frac {g(k_0r)}{k_0r} \right] ,
\end{equation}
and the near- and far-zone approximations yield
\begin{equation}\label{enshiftpar2}
\Delta E_\parallel (r) \simeq -\beta \mathcal{E} \mathcal{E}' \frac 1{k_0^2} \times
\left\{
\begin{array}{cc}
& \frac \pi{r^3} \, \, \, \, \, \, \mbox{for $k_0r \ll 1$} \\
& \\
& \frac 4{k_0r^4} \, \, \, \,  \mbox{for $k_0r \gg 1$} .
\end{array}
\right.
\end{equation}

Both \eqref{enshiftperp2} and \eqref{enshiftpar2} show that the field-assisted component of the dispersion interaction decreases as $r^{-3}$ in the near zone, and as $r^{-4}$ in the far zone.
The field-assisted component can now be compared with the unperturbed dispersion energy between ground-state atoms, given by \cite{CP98,Buhmann1,Salam10,Salam08}
\begin{equation}\label{vdW}
\Delta E_{vdW} (r) \simeq
\left\{
\begin{array}{cc}
& - \frac 3{64\pi^2 \epsilon_0^2} \bar{E} \alpha^2 \frac 1{r^6} \, \, \, \, \, \, \mbox{for $k_0r \ll 1$}, \\
& \\
& -\frac {23 \hbar c}{64 \pi^3 \epsilon_0^2} \alpha^2 \frac 1{r^7} \, \, \, \, \, \, \, \mbox{for $k_0r \gg 1$} ,
\end{array}
\right.
\end{equation}
where $\bar{E}$ is an average atomic excitation energy, defined as $\bar{E}=E_AE_B/(E_A+E_B)$, $E_A$ and $E_B$ being the most relevant excitation energy of atoms $A$ and $B$ (from the ground to the first excited level), and $\alpha$ their static polarizability. When the external field is present, the complete fourth-order interaction energy is
$\Delta E_{vdW} (r) + \Delta E_{\perp (\parallel )} (r)$.
\begin{figure}[h]
\centering
\includegraphics[width=0.45\textwidth]{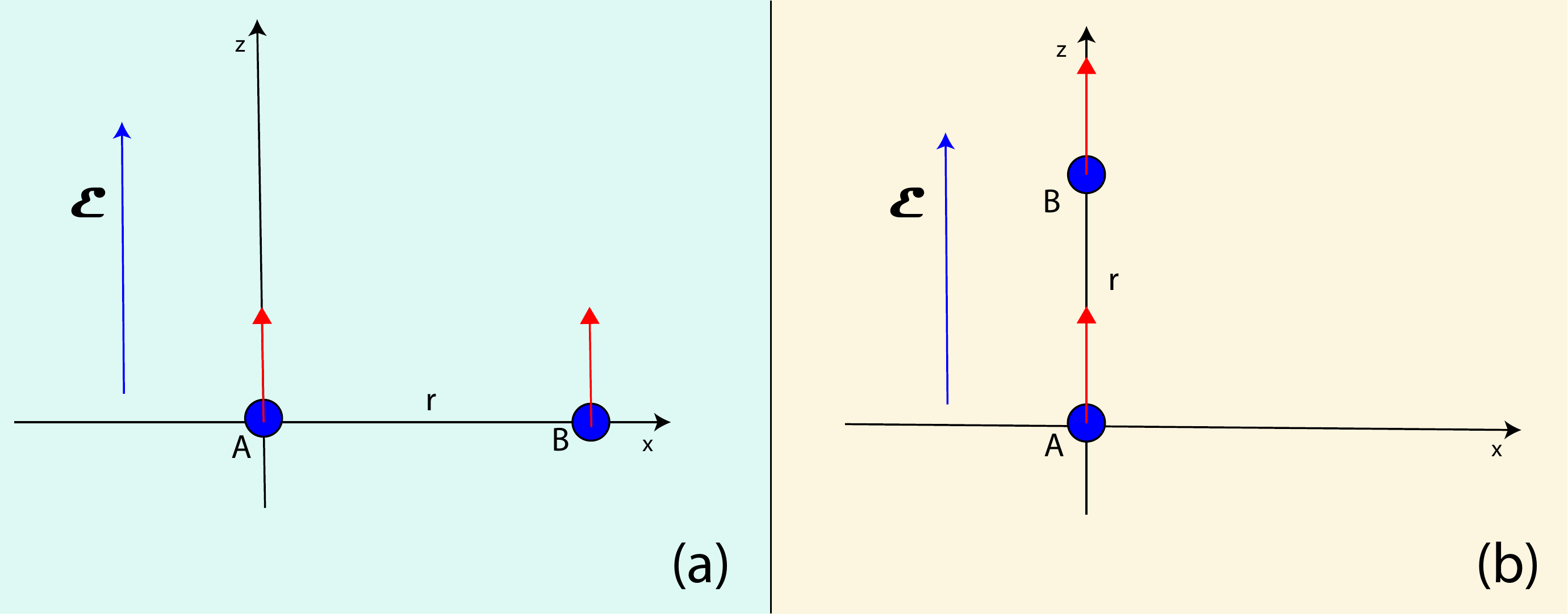}
\caption{Two hydrogen atoms in free space in the presence of an external static electric field $\mathcal{E}$ along the $z$ direction: (a) atoms aligned perpendicularly to the direction of the external field; (b) atoms aligned in the same direction of the external field. The red arrows indicate the direction of the induced dipole moments.}
\label{fig:system}
\end{figure}

The field-dependent component of the dispersion interaction scales with the distance quite slower than the unperturbed interaction; indeed, its scaling is the same as that of the attractive atom-wall Casimir-Polder interaction, $r^{-3}$ in the near zone, and $r^{-4}$ in the far zone,
as a comparison with the expressions in the Casimir-Polder original paper shows \cite{CasimirPolder48}.
Its sign, determining its attractive or repulsive character, can however be different, depending on the geometric configuration and the relative orientation of the external electric fields.

Figure \ref{fig:system} shows, for $\mathbfcal{E} = \mathbfcal{E}'$, the two configurations considered, perpendicular and parallel. The arrows on the atoms indicate the direction of the induced dipole moments. Our results \eqref{enshiftperp1},\eqref{enshiftperp2},\eqref{enshiftpar1} and \eqref{enshiftpar2} show that, when $\mathbfcal{E}$ and $\mathbfcal{E}'$ are parallel, the change in the dispersion interaction energy due to the external field is positive in the perpendicular configuration (yielding a repulsive contribution), while it is negative in the parallel configuration (attractive contribution).
This can be understood by a transparent physical picture in terms of the interaction between the induced dipole moments: in the perpendicular case [Figure \ref{fig:system}(a)], the dipole-dipole interaction gives a repulsive force (the interaction energy of one induced dipole in the field generated by the other dipole, is positive), whereas in the parallel case [Figure \ref{fig:system}(b)] it yields an attractive force (the interaction energy is negative). The situation is reversed if the electric fields on the two atoms have opposite directions, because in this case $\mathcal{E}$ and $\mathcal{E}'$ have an opposite sign, and the induced dipoles are opposite. On the other hand, the unperturbed interaction \eqref{vdW} between ground-state atoms is always attractive. Thus our findings show that, exploiting external static electric fields, we can modify the atomic dispersion interaction, and even turn it from attractive to repulsive for sufficiently intense external electric fields.

A similar physical picture holds for the (ground-state) atom-plate Casimir-Polder interaction, in terms of the interaction between the instantaneous atomic dipole and its image beyond the plate, that are correlated: the components parallel to the surface point in opposite directions, while the perpendicular component points to the same direction. For an isotropic atom, this eventually yields the attractive atom-plate Casimir-Polder potential, due to exchange of one virtual photon, and behaving as $r^{-3}$ (near zone) and $r^{-4}$ (far zone) \cite{Barnet00,Antezza14}. There is some analogy also with the resonance interaction between two entangled atoms, one excited and the other in the ground state \cite{Chibisov72,Incardone14,Passante18}, where the unperturbed atomic dipole moments are correlated too, and a one-photon exchange is involved. The resonance interaction has however a different scaling with the distance, due to the possibility of a real photon exchange.

We now compare numerically the field-dependent contribution which we have obtained with the unperturbed dispersion interaction, when the interatomic distance is $r=10^{-6} \, \mbox{m}$. 
Since $k_0 \simeq 1.03 \times10^8 \, {\mbox{m}}^{-1}$, we have $k_0r \gg 1$ (far zone). From \eqref{enshiftperp2} and \eqref{enshiftpar2}, we obtain
\begin{equation}\label{stima1}
\Delta E_\perp = -\Delta E_\parallel \simeq 1.7 \times 10^{-36} \, \mathcal{E} \mathcal{E}' \, \, {\rm eV/(V/m)^2} .
\end{equation}

At the same distance, the unperturbed dispersion energy \eqref{vdW} is
\begin{equation}\label{stima2}
\Delta E_{vdW} \simeq -7.8 \times 10^{-27} \, \mbox{eV}
\end{equation}
[for consistency, in the evaluation of \eqref{stima2}, we have included only contributions to the polarizability from $n=2$ states].

A comparison of \eqref{stima1} with \eqref{stima2} shows that, at the distance considered, $\lvert \Delta E_\perp \rvert$ and $\lvert \Delta E_\parallel \rvert$ become comparable with $\lvert \Delta E_{vdW} \rvert$  for a field strength of the order of $\mathcal{E} = \mathcal{E}' \simeq 10^5 \, \mbox{V/m}$. Such an intensity is well within the reach of static fields currently obtained in the laboratory
\cite{Bailey95,Maron86,GonzalezFerez09}, and within, even by several orders of magnitude, the strength limit imposed by our perturbative treatment of the external field (Stark shifts much smaller than atomic transition energies). With the same strength of the electric field, and at larger interatomic distances, the field-modified interaction can exceed the unperturbed interaction energy by several orders of magnitude. All this indicates a realistic possibility to observe the new effects which we have calculated. At shorter distances, our results \eqref{enshiftperp1},\eqref{enshiftperp2},\eqref{enshiftpar1},\eqref{enshiftpar2} show that higher field intensities are required for making the field-mediated contribution comparable to the unperturbed one: $\sim 10^6 \,\,  \mbox{V/m}$ for $r \sim 100 \, \mbox{nm}$, and $\sim 10^8 \, \, \mbox{V/m}$ for $r \sim 10 \, \mbox{nm}$ (both in the near zone). Submicrometrical distances, as those which we are considering, are a typical distance between the parts of MEMS and NEMS, where Casimir dispersion interactions become relevant \cite{Chan01}.
In the configurations yielding a repulsive field-assisted component of the dispersion force, the external fields and the interatomic distance can be appropriately calibrated to make the total dispersion interatomic force vanish (this equilibrium distance, however, is an unstable point). As mentioned, all this can also have relevance in applications where Casimir forces are important \cite{Chan01,Bostrom12}.

We now compare our results with previous ones for dispersion interactions in the presence of a radiation field, in particular Ref. \cite{Thiru80}, where the interaction between two nonpolar molecules in the presence of an external electromagnetic field in a Fock state is considered. Notwithstanding some analogy, this case is fundamentally different from ours, because a photon in a Fock state, contrarily to a static field, does not induce a permanent dipole moment in the atoms (the average value of the electric field in a Fock state is zero); thus, our results cannot be obtained from those in \cite{Thiru80} even in the zero-frequency limit. This is confirmed by the different distance scaling of the interaction energy, in the retarded Casimir-Polder regime: $r^{-2}$ or $r^{-3}$ with space oscillations (similarly to the case of excited atoms) in \cite{Thiru80}, depending on the polarization of the photon, whereas in our case we find a $r^{-4}$ monotonic behavior, allowing to reverse the sign of the dispersion force over a large portion of space. Also, the physical origin of the change of the distance scaling is very different in the two cases: possibility of atomic excitation by absorption of the external photon in \cite{Thiru80}, and permanent correlated dipole moments in the ground-state atoms, induced by the static field, in our case.
Our results are also coherent with the experimental and theoretical results reported in \cite{Jahangiri19}, where the near-zone van der Waals interaction between Rydberg atoms in $S$ states under the action of an electric field is considered, using a phenomenological model where a $r^{-3}$ electrostatic dipole-dipole interaction between the atoms is added, obtaining good agreement between theoretical predictions and experiments (they consider only the near zone); a similar near-zone model was used in \cite{You98}. The effect of an electric field on the van der Waals dipole of a molecule pair \cite{Hunt85} has been also considered.

\textit{Conclusions}.
In this Letter we have considered the dispersion interaction (van der Waals and Casimir-Polder) between two ground-state hydrogen atoms subjected to an external static electric field, and shown that the external field can be exploited to strongly modify the intensity, distance dependence, and character (attractive or repulsive) of the dispersion force. We have estimated, at typical distances and for field intensities such that can be treated by perturbation theory, the value of the field-modified component of the interaction in relevant configurations, and compared it with the dispersion interaction for unperturbed atoms. Our new findings show that, with a strength of the external field currently attainable in the laboratory, the force can be strikingly tailored exploiting the external field, with the possibility of obtaining a significant increase or decrease, or reversing it from attractive to repulsive. A significant enhancement of the force could be of striking importance for a direct detection of such interactions, in particular in the retarded Casimir-Polder regime \cite{Przybytek12,Beguin13}. As shown by Eqs. \eqref{enshiftperp2} and \eqref{enshiftpar2}, a repulsive dispersion interaction can be achieved, with appropriate field strengths, when the electric fields are parallel to each other and the atoms are aligned in a direction perpendicular to the external field, or when the external electric fields are antiparallel to each other and the atoms are aligned parallel to the external fields. Our work has been concerned with ground-state atoms; a possible future extension will consider excited atoms, where the exchanged photon can be real, for which there has been a renewed interest in recent literature \cite{Donaire15,Milonni15, Barcellona16,Jentschura17}.

\begin{acknowledgments}
The authors gratefully acknowledge financial support from the Julian Schwinger Foundation and MIUR.
\end{acknowledgments}

\end{document}